\documentclass[proof]{WileyASNA-v1}
\usepackage[utf8]{inputenc}
\usepackage{booktabs}
\usepackage{caption}
\usepackage{graphicx}
\usepackage{verbatim}
\usepackage{bm}
\makeatletter

\newcommand{\Rmnum}[1]{\expandafter\@slowromancap\romannumeral #1@}
\makeatother
\articletype{Article Type}%

\begin{document}

\title{A short review of the pulsar magnetic inclination angles (\Rmnum{2})}

\author[1,3]{Biao-Peng Li}

\author[1,3]{Wen-Qi Ma}

\author[1,2,3]{Zhi-Fu Gao*}

\authormark{Biao-Peng Li}

\address[1]{Xinjiang Astronomical Observatory, CAS,150,
Science 1-Street, Urumqi, Xinjiang, 830011, China}

\address[2]{Key Laboratory of Radio Astronomy, Chinese Academy of Sciences,
West Beijing Road, Nanjing, 210008, China}

\address[3]{University of Chinese Academy of Sciences, No.19, Yuquan Road, Beijing, 100049, China}

\corres{*Zhi-Fu Gao. Xinjiang Astronomical Observatory, CAS,150, Science
1-Street, Urumqi, Xinjiang, 830011, China.\\
\email{zhifugao@xao.ac.cn}}

\abstract{The pulsar magnetic inclination angle is a key parameter for pulsar physics. It influences the observable
properties of pulsars, such as the pulse beam width, braking index, polarisation, and emission
geometry. In this study, we give a brief overview of the current state of knowledge and research
on this parameter and its implications for the internal physics of pulsars. We use the observed
pulsar data of magnetic inclination angle and braking index to constrain the star's number of precession
cycles, $\xi$, which reflects the interaction between superfluid neutrons and other particles inside a neutron star\,(NS). We apply the method proposed by Cheng et al. (2019) to analyse the data of PSR J2013+3845 
and obtain the constraints for $\xi$ ranging from $2.393\times 10^{5}$ to $1.268\times10^{6}$. And further analysis suggests that the internal magnetic 
field structure of PSR J2013+3845 is likely dominated by toroidal component. This study may help us understand 
the process of internal viscous dissipation and the related evolution of the inclination angles of pulsars, and 
may have important implications for the study of continuous gravitational wave emissions from NS.}


\fundingInfo{ The Natural Science Foundation of Xinjiang Uygur Autonomous Region, China (Grant No. 2022D01A155), the Major Science and Technology Program of Xinjiang Uygur Autonomous Region (Grant No. 2022A03013-1), and NSFC (Grant No. 12288102). }

\maketitle

\section{Introduction}\label{sec1}
Pulsars are highly magnetized and rapidly rotating neutron stars, providing us an opportunity to explore physics under extreme conditions~\citep{Deng2020, Wen2021, Deng2021, Shan2022, Shan2023}. The models of spin-down, and magneto-thermal evolutions of pulsars have been used by many authors~\citep{Yuan2017, Yan2018, Yan2020, Kou2019, Wang2019, Wen2022} as a way of probing the internal composition of these objects. Such studies rely on the fact that the physical quantities relevant for the rotation instability and energy loss of pulsars~\citep{Gao2017, Gao2021, Liu2017, Liu2019}. An important physical quantity is the magnetic inclination angle, which is the angle between the rotation axis and the magnetic axis of a pulsar. This physical quantity determines the emission shape and intensity of a pulsar, as well as its spin-down rate and braking index. Therefore, studying the inclination angles and their evolutions of pulsars can help us understand the origin, evolution, and diversity of these fascinating objects~\citep{Zhou2021, YanZhen2018, Wen2020, Deng2020}, and can help us probe the internal structure and dynamics of neutron stars, and the extreme physics of matter and fields under high density and strong gravity.

Recently, we provided a brief overview of the current research on the evolution of the pulsar magnetic inclination angle~\citep{Li2023}. Firstly, the paper introduces the basic concepts and definitions of the pulsar magnetic inclination angle, and reviews the main mechanisms that can cause the angle to change over time, such as the magnetic dipole radiation, plasma-filled magnetosphere, magnetic field decay, and gravitational wave emission. Then the paper presents some of the recent results and progress on the inclination angle of pulsars, uses the alignment rotator model in a vacuum to investigate the inclination angle change rates for these pulsars, and compare them with the values obtained from other methods for three representative cases: J0157+6212, J1743-3150, and J1857+0526. At last the paper gives a brief summary and outlook. The paper concludes that the inclination angle of pulsars may not be constant over time, as various physical processes can cause it to change. For example, magnetic field decay, accretion from a companion star, or glitches (sudden changes in the pulsar’s rotation)~\citep{Yuan2010, Kou2018, Zhou2022} can alter the magnetic inclination angle of pulsars. These processes can also affect the pulsar’s emission and spin-down behavior, making it more complex and diverse.

The reminder of this paper is organized as follows. In Section 2 we give a brief overview of inclination angle $\chi$ of pulsars from five aspects: inclination angle variation; measurements of $\chi$ of pulsars, possible observable properties in pulsars caused by variations in $\chi$, connection between inclination angle and braking index, and connection between the inclination angle and the number of precession cycles. In Section 3, we obtain a constraint for the number of precession cycles $\xi$ for PSR J2013+3845 with the measured values of its braking index and inclination angle, and Section 4 gives summary and outlook.

\section{Overview of inclination angle }\label{sec2}
\subsection{Inclination angle variations}\label{sec2.1}
As we know, the magnetic inclination angle of a pulsar is an important parameter in the radiation geometry and plays a decisive role in determining the structure of the magnetosphere and radiation beam. The magnetic inclination angle is a valuable parameter for studying pulsar radiation, dynamics, and evolution. Various models have been proposed to describe the process of magnetic inclination evolution. ~\cite{Michel1970} considered the case of a pulsar rotating in vacuum. By integrating the Maxwell stress exerted at the surface of the rotator, they obtained the torque equations acting on the rotating magnetic dipole. They found that the evolution of $\chi$ in this vacuum model decreases exponentially with time. However, realistic pulsars have a co-rotating magnetosphere filled with plasma. Taking into account the plasma effects, ~\cite{Philippov2014} analyzed the results of magnetohydrodynamic simulations of pulsar magnetospheres and found that the evolution of $\chi$ in plasma-filled pulsars decreases with time following a power-law form which is $\chi\propto(t/\tau)^{-1/2}$, where $\tau \sim $ spin-down time-scale. According to ~\cite{Beskin1993}, a surface current exists within the polar cap of a pulsar. This surface current closes the longitudinal current in the magnetosphere and the return current that flows along the separatrix between the open and closed field lines regions. Some part of their path must cross the magnetic field. Consequently, a Lorentz force is induced by these currents in the polar cap region, resulting in torques acting on the crust of NS. This current loss model (BGI model)~\citep{Beskin1993} predicts that $\chi$ will increase with time. In addition, there are some models which give different evolution forms, such as precession model~\citep{Zanazzi2015, Lander2018} and two-dipole model~\citep{Hamil2016}. In particular, for pulsars in accretion state, the acceleration torque can affect the evolution of $\chi$~\citep{Biryukov2021}. Using the binary evolution code MESA, ~\cite{Yang2023} simulate the evolution of $\chi$ of accreting NS in intermediate/low X-ray binaries and show that the evolution of $\chi$ depends not only on the initial parameters of the binary systems, but also on the mass transfer history and the efficiency of pulsar's rotational energy loss. And in statistical analysis, it is usually concluded that the magnetic inclination decreases with time~\citep{Rankin1990, Tauris1998, Faucher2006, Weltevrede2008, Young2010, Maciesiak2011, Gullon2014, Kenko2023}. But ~\cite{Beskin1993} showed that, given the dependence of $\chi$ on the extinction line, even if $\chi$ of a single pulsar increases over time, the observed mean value of $\langle\chi\rangle(\tau_d)$ can decrease, where $\tau_{d}=P/\dot{P}$ is the dynamical age.
\subsection{Measurements of \texorpdfstring{$\chi$}{chi} of pulsars}\label{sec2.2}
There are several methods to determine the magnetic inclination angle $\chi$. \cite{Radhakrishnan1969} proposed the rotating vector model (RVM) to explain the S-shaped polarization position angle (PPA) curve obtained from polarimetric observations of pulsars. The plane of linearly polarised emission is determined by the direction of the magnetic field at the point of emission. When the beam sweeps across the observer, the projected direction rotates with NS and the measured PPA varies at the outer wings of the profiles and change rapidly at the profile centre. The most widely used method for obtaining the magnetic inclination angle is to fit the polarization position angle of pulsars using the rotating vector model. Recently \cite{Wang2023} utilized this method to fit the polarization observation data from FAST and obtained geometric parameters, including the magnetic inclination angle, for 190 pulsars. However this classical rotation vector model has theoretical limitations. There are several effects that can cause the PPA curve to deviate from the prediction of this model, including the effects of aberration, the delay caused by height differences, etc. \cite{Blaskiewicz1991} presented a relativistic model of pulsar polarization which involves radio emission from the open field line region at radii well within the light cylinder. In addition, \cite{Tong2021} calculated the modification of the rotating vector model in the case of magnetars which have twisted magnetic field compared with normal pulsars. And \cite{Qiu2023} studied the RVM in the presence of a multipole magnetic field in pulsars and magnetars. He found that the expression for the PPA is unchanged and only $\chi$ and phase constant $\phi_{0}$ will change. 

\cite{Rankin1990} discovered a simple mathematical 
relationship between the pulse width $W_{\mathrm{core}}$ and the rotation period and magnetic inclination
\begin{eqnarray}
W_{\mathrm{core}}=2^{\circ}.45P^{-1/2}/\sin{\chi}.
\label{eq1}
\end{eqnarray}
~\cite{Rankin1990} used Equation (\ref{eq1}) to determine $\chi$ of about 110 pulsars. ~\cite{Maciesiak2011MN} performed a statistical analysis of half-power pulse-widths of the core components in average pulsar profiles and found that this Rankin's method of estimation of $\chi$ in pulsars with core components is quite good an approximation, except for very small angles in the almost aligned rotators. For the gamma-ray pulsar, we can use the gammy-ray radiation model to fit the pulse profile to get the magnetic inclination~\citep{Johnson2014}. However, this method has great model dependence. Besides, there are some methods to determine the $\chi$ for special pulsars~\citep{Smits2007, Manchester2010, desvignes2012}. However, measuring the magnetic inclination angle of pulsars is not easy, as it requires accurate modeling of the pulsar’s emission geometry and timing properties \citep{Wang2005, Wang2008, WenZG2016, Wen2016, WenZG2020}.
\subsection{Observable properties caused by variations in \texorpdfstring{$\chi$}{chi}}\label{sec2.3}
As mentioned above, we have summarized the main theoretical models and observational methods for studying the inclination angle of pulsars,
discussed some of the latest results and challenges in this field, and also pointed out that variation in $\chi$ may affect the observable 
properties of pulsars, such as the pulse beam width, the braking index, and the polarimetric emission \citep{Li2023}.

\cite{Arzamasskiy2017} found that statistical study of interpulse pulsar (IP) is the key to solve the problem of magnetic inclination evolution. Because IP is generally believed to be a pulsar with $\chi\approx0^{\circ}~\text{or} ~90^{\circ}$~\citep{Manchester1977, Lyne1998}, it can provide additional information about $\chi$. However, due to the large uncertainties in the distribution of pulsars over their initial inclination angles and initial spin periods, both the BGI model and the plasma-filled model can explain the number of IP observed.

Because the magnetic inclination evolves very slowly, with a timescale of about $10^{6}-10^{7}$ years~\citep{Tauris1998, Young2010}, it is difficult to measure the rate of change of $\chi$. At present, only the Crab pulsar has given the inclination change rate of $\dot{\chi}=0.62^\circ\pm0.03^\circ/100\text{yrs}$ by measuring the deviation of the interpulse from the main pulse~\citep{Lyne2013}. However, \cite{Beskin2018} do not believe that this value is credible because of some model assumptions made in the paper, such as the gamma-ray pulse profile changed with time in the same way as in the radio range. We hope that more accurate and longer time observations for pulsars and the perfect theoretical model will be proposed in the future.
\subsection{Connection between \texorpdfstring{$\chi$}{chi} and braking index}\label{sec2.4}
The braking index, $n$, of a pulsar describes the rate at which it loses rotational energy. It is an observable quantity, which can be expressed in terms of the pulsar's rotation angular velocity $\Omega$, rotation frequency $\nu$ or rotation period $P$:
\begin{eqnarray}
    n=\frac{\Omega\ddot{\Omega}}{\dot{\Omega}^2}=\frac{\nu\ddot{\nu}}{\dot{\nu}^2}=2-\frac{P\ddot{P}}{\dot{P}^2}.
    \label{eq2}
\end{eqnarray}

In addition to magnetic dipole radiation (MDR), pulsar wind and gravitational wave emission (GWE) are also considered as pulsar braking mechanisms. These radiation mechanisms gives the braking index as $n=3, 1$ and 5, respectively~\citep{Ostriker1969}. Other braking mechanisms have been investigated, such as quantum vacuum friction~\citep{Dupays2008}, precession~\citep{Dall'Osso2009}, etc. These above processes are often combined to explain the observed braking indices~\citep{Gao2016, Gao2017, Gao2021}.

The MDR gives the braking index of $n=3$ on the basis that the moment of inertia $I$, magnetic moment $\mu=BR^{3}$ ($B$ is the dipole magnetic field in the pole of the star, and $R$ is the stellar radius) and magnetic inclination angle $\chi$ are constants. However, $I$, $\mu$ and $\chi$ any change, any change in these parameters will result in a deviation of $n$ from $3$. For example, from the following expression
\begin{eqnarray}
    n=3+2\frac{\nu}{\dot{\nu}}\left(\frac{\dot{I}}{I}+2\frac{\dot{\chi}}{\tan\chi}+\frac{\dot{B}}{B}\right),
    \label{eq3}
\end{eqnarray}
one can see that if $\chi$ increases, $n$ will be less than 3 and vice versus, assuming that both $B$ and $I$ are constant.

Studying the braking index and its evolution is necessary for several reasons. Firstly, it can help us test different theories of pulsar emission and magnetism, and constrain the physical parameters of pulsars. Secondly, it can help us explore the connection between different types of pulsars, such as normal pulsars, millisecond pulsars, and magnetars~\citep{Mereghetti2008, Gao2016, Gao2021, Huang2021, Huang2023}, and understand their evolutionary pathways. Thirdly, it can help us use pulsars as tools to study other astrophysical phenomena, such as gravitational waves, binary evolution, and interstellar medium\citep{Stella2005, Zhou2021, Deng2021, Hu2023}.
\subsection{Connection between \texorpdfstring{$\chi$}{chi} and the number of precession cycles}\label{sec2.5}
Generally speaking, for a highly magnetized neutron star, the axis of symmetry of its magnetic deformation does not coincide with the axis of rotation, so its angular velocity vector will precess freely on the symmetry axis of its magnetic deformation with a period of $P_{\mathrm{prec}}\simeq P/\epsilon_{B}$, where $\epsilon_{B}$ is the ellipticity of the magnetic deformation.   The rotation of the star causes periodic changes in the quadrupole moment of the mass, which in turn produces GWE~\citep{Jones2002}. The free precession process is damped by the viscous dissipation inside neutron star, resulting in a change of magnetic inclination angle~\citep{Dall'Osso2009}. The change rate of $\chi$ in this process is related to the number of precession cycles $\xi$~\citep{Jones1976}. \cite{Cutler2002} defines $\xi$ as the dissipative timescale divided by the precession period, i.e., $\xi\equiv\tau_{\mathrm{DIS}}/P_{\mathrm{prec}}$. At present, the internal dissipation process of a neutron star is not very clear, so only a rough estimate of $\xi$ can be given. For example, \cite{Alpar1988} studied the core-shell coupling caused by the scattering of electrons in the neutron vortex, and obtained $\xi\approx10^{2}-10^{4}$. The number of precession cycles represents specific viscous mechanisms through which the precession energy of neutron star is dissipated during precession~\citep{Cheng2019}. In addition, this parameter is extremely important when discussing continuous GWE from a pulsar~\citep{Gualtieri2011}.
\begin{table*}[ht]
\centering
\caption{The spin parameters of PSR J2013+3845, including the spin frequency, its first derivative, second derivative and inclination angle. References: [1] \cite{Hobbs2010}; [2]\cite{Wang2023}.
\label{tab1}}
\vspace{-1 em}
\scalebox{1.5}{
\begin{tabular}{lccccccccc}
\bottomrule
Pulsar & $P$ & $\dot{P}$ & $n$ & Ref &$\chi_{\text{obs}}$ &Ref & $\tau_{\text{c}}$  \\
       & (s) & $(10^{-15} \text{s}/\text{s})$  &  &     & ($^{\circ}$)     &      &($10^{5}\text{yrs}$)  \\
\midrule
J2013+3845  & 0.230 & 8.851    & 2.34(17) & [1]  & 52 & [2] & 4.123      \\
\bottomrule
\end{tabular}}
\vspace{-1.3em}
\end{table*}

A pulsar has strong internal magnetic fields, which can cause the pulsar to deform significantly which can be represent by ellipticity $\epsilon_{B}$. It is usually difined as $\epsilon_{B}=\Delta I/I_{3}$, where $\Delta I\equiv I_{3}-I_{1}$ is the distortion being sourced by magnetic strains, and $I_{1}, I_{2}=I_{1}, I_{3}$ is the principal moments of
inertia. If the magnetic field is dominated by toloidal component, the NS has a problate shape ($\epsilon_{B}<0$). And a poloidal field, by contrast, deforms a star in an oblate ($\epsilon_{B}>0$) shape~\citep{Lander2018}. For a prolate pulsar, these internal processes may orthogonalize the star's magnetic symmetry axis with the angular momentum vector through viscous dissipation~\citep{DallOsso2017}. At this time, the rotational energy of a pulsar is minimum and the star is in a geometrical state where the quadrupole moment changes the most, thus improving the efficiency of GWE.~\citep{Stella2005}.

\section{Constraining \texorpdfstring{$\xi$}{xi} of PSR J2013+3845}\label{sec3}
Here we give a constraint for the number of precession cycles $\xi$ for PSR J2013+3845 with the measured values of its braking index $n$ and inclination angle $\chi$. This section includes two parts. In the former we introduce a theoretical model proposed by \cite{Cheng2019}, in which free precession process is damped by viscous dissipation inside NS, resulting in a change of $\chi$. In the latter, the data of $n$ and $\chi$ of PSR J2013+3845 is analyzed and a constraint for the number of precession cycles $\xi$ is given.
\subsection{Theoretical model}\label{sec3.1}
As discussed above, the rotation of an isolated pulsar will be slowed down by MDR and GWE, and the strong magnetic fields will cause a magnetic deformation of the star, which leads to precession. And the free precession of pulsar is damped by viscous dissipation inside the star. As the pulsar spins down, the magnetic inclination also changes Meanwhile, the crust magnetic fields are undergoing a decay under the action of Hall drift and Ohmic dissipation. Considering these above effects, ~\cite{Cheng2019} described the evolution of $\chi$ of a pulsar by the following expression
\begin{eqnarray}
    \left.\dot{\chi}=\left\{\begin{aligned}-\frac{2G}{5c^5}I\epsilon_\mathrm{B}^2\Omega^4\sin\chi\cos\chi(15\sin^2\chi+1)-\frac{\epsilon_\mathrm{B}}{\xi P}\mathrm{tan}\chi\\-\frac{kB_\mathrm{d}^2R^6\Omega^2}{Ic^3}\sin\chi\cos\chi,\mathrm{~for~\epsilon_\mathrm{B}>0}\\-\frac{2G}{5c^5}I\epsilon_\mathrm{B}^2\Omega^4\sin\chi\cos\chi^2(15\sin^2\chi+1)-\frac{\epsilon_\mathrm{B}}{\xi P}\mathrm{cot}\chi\\-\frac{kB_\mathrm{d}^2R^6\Omega^2}{Ic^3}\sin\chi\cos\chi,\mathrm{~for~\epsilon_\mathrm{B}<0},\end{aligned}\right.\right.
    \label{eq4}
\end{eqnarray}
where $G$ is the gravitational constant, $c$ the speed of light. The moment of inertia $I$ is typically taken to be $10^{45}\operatorname{g}\operatorname{cm}^2$, and the stellar radius $R=10\text{km}$. Here $k=1/6$ is taken as a constant associated with MDR. The first and third terms of the Equation (\ref{eq4}) represent the decrease of $\chi$ caused by GWE and MDR, respectively, and the second term represents the evolution of $\chi$ caused by free precession, which depends on the sign of $\epsilon_\mathrm{B}$.

Assuming that there is a co-rotating magnetosphere outside a pulsar, its braking index under this model is given as \citep{Cheng2019}
\begin{eqnarray}
    \begin{aligned}n&=3-\frac{2P}{\dot{P}}\biggl\{\frac{\dot{B}_{\mathrm{d}}}{B_{\mathrm{d}}}+\dot{\chi}\sin\chi\cos\chi\biggl[\frac{1}{1
    +\sin^2\chi}\\&+\frac{1+30\sin^2\chi}{\eta\sin^2\chi(1+15\sin^2\chi)}\biggr]\biggr\},\end{aligned}
    \label{eq5}
\end{eqnarray}
where $\eta=5kc^2B_{\mathrm{d}}^2R^6(1+\sin^{2}\chi)/[2G\epsilon_{\mathrm{B}}^{2}I^{2}\Omega^{2}(1+15\sin^{2}\chi)\allowbreak\sin^{2}\chi]$ is the ratio of MDR spin-down to GWE spin-down rates.
\begin{figure}[t]
\vspace{-0.3 cm}
\centering
\includegraphics[width=3.8in, height=3.2in]{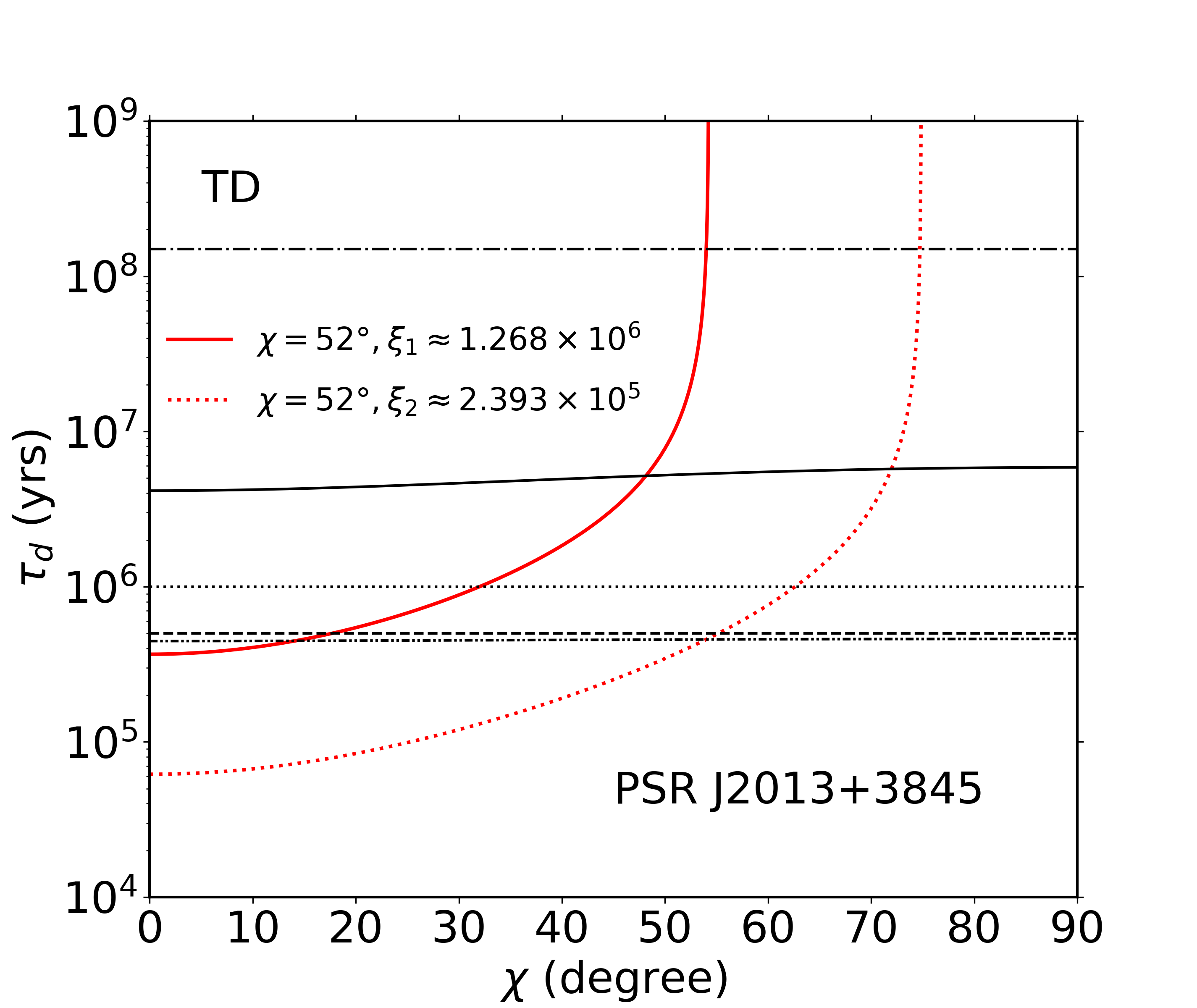}
\caption{Evolution curve of dipole field decay timescale $\tau_{\mathrm{D}}$ with magnetic inclination angle $\chi$ of PSR J2013+3845. The red lines is the constrains on $\xi$ after taking into account the observation of $\chi$. The black curves are obtained based on the theory of magnetic field decay. See Section \ref{sec3.1} for more details.}
\label{fig1}
\vspace{-0.5 cm}
\end{figure}
The ellipticity $\epsilon_\mathrm{B}$ depends on the internal magnetic field energy, the internal magnetic field structure and the equation of state. For the ellipticity of a pulsar with a oblate shape, we take $\epsilon_\mathrm{B}=3.4\times10^{-7} (\frac{B_{\text{d}}}{10^{13}\text{G}})$~\cite{Lander2013}. As to the ellipticity of a prolate pulsar, we take $\epsilon_\mathrm{B}\approx -B_{\text{d}}\times10^{-20}$~\citep{Akgun2008}.

For the dipole magnetic field decay, a simple exponential form is adopted, $\dot{B_{\mathrm{d}}}=-B_{\mathrm{d}}/\tau_{\mathrm{D}}$, where $\tau_{\mathrm{D}}$ is the timescale for magnetic field decay~\citep{Pons2007P, DallOsso2012}. Let us consider the following three scenarios. When Hall drift dominates the decay process, then $\tau_{\mathrm{D}}=\tau_{\mathrm{H}}\simeq1.2\times10^{4}(B_{\mathrm{d}}/10^{15}\mathrm{G})^{-1}\mathrm{yrs}$~\citep{Cumming2004}, and $B_{\mathrm{d}}=\left[-\frac{\dot{\omega}Ic^{3}}{k\omega^{3}R^{6}(1+\sin^{2}\chi)}\right]^{1/2}$, so the field decay timescale can be expressed as a function of $\chi$, which is represented by a solid black line in Figure \ref{fig1}. And if Ohmic dissipation dominates the decay process,  $\tau_{\mathrm{D}}=\tau_{\mathrm{O}}=5\times10^{5},10^{6},1.5\times10^{8}$ yrs~\citep{Pons2007P, Bransgrove2018}. They are represented by black dashed, dotted, and dashed-dotted lines in Figure \ref{fig1}. The third scenario is to consider that the dipole field decays due to the combined effects of Hall drift and Ohmic dissipation, then $\tau_{\mathrm{D}}=1/[1/\tau_{\mathrm{H}}+1/\tau_{\mathrm{O}}]$. In this case, substitute $\tau_{\mathrm{O}}=5\times10^{5}\mathrm{yrs}$ into the above formula to get the minimum $\tau_{\mathrm{D}}-\chi$ curve~\citep{Cheng2019}. It is represented by a black dashed-dotted-dotted line in the Figure \ref{fig1}.

\begin{figure}[t]
\vspace{-0.3 cm}
\centering
\includegraphics[width=3.8in, height=3.2in]{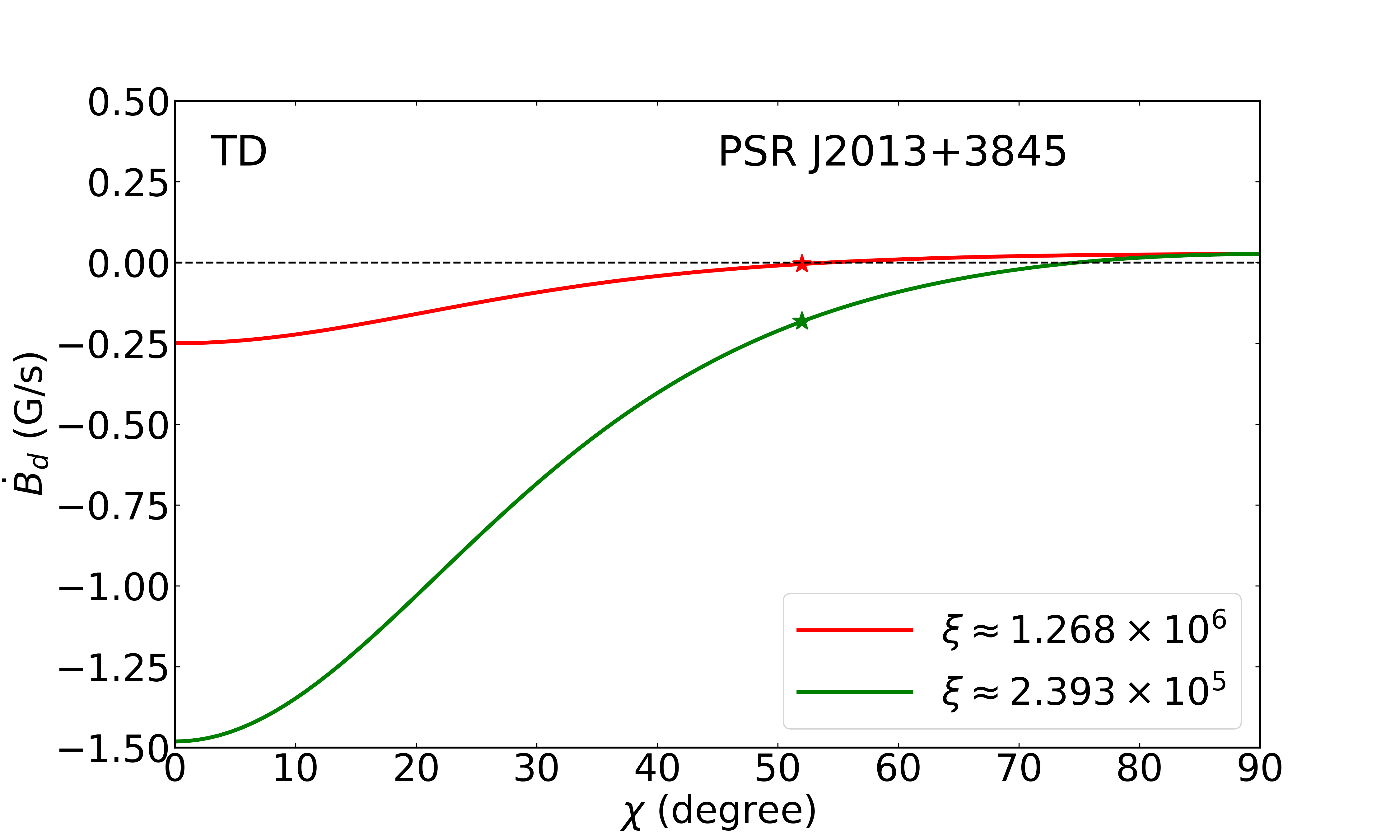}
\caption{Evolution curve of the dipole field decay rate $\dot{B_{\mathrm{d}}}$ with magnetic inclination angle $\chi$ for PSR J2013+3845. It is assumed that the internal magnetic field of NS is dominated by the toroidal component.}
\label{fig2}
\vspace{-0.5 cm}
\end{figure}
\subsection{Sample of PSR J2013+3845}\label{sec3.2}
We select PSR J2013+3845 as a sample for  analysis, ignoring the errors. For this pulsar, we select the timing observation data in ~\cite{Hobbs2010}, whose observation time span is 18.5 years. The vale of magnetic inclination $\chi$ is taken from ~\cite{Wang2023}, where the RVM was used in fitting the latest FAST polarization data to obtain $\chi$. The parameters of J2013+3845 are shown in Table \ref{tab1}.

Following the ~\cite{Hu2023}, for PSR J2013+3845 with a braking index $n < 3$, we assume that the internal magnetic field is dominated by the toroidal components. We substituted the observed values of $P$, $\dot{P}$, $n$ into Equation (\ref{eq4}) and Equation (\ref{eq5}) to obtain the evolution curve of $\tau_{\mathrm{D}}$ with $\chi$ at different values of $\xi$. By causing the $\tau_{\mathrm{D}}-\chi$ curve obtained from the timing observation to intersect with the upper bound of the magnetic decay theory at the observed magnetic inclination, we obtain the upper limit of $\xi$ is $\xi\approx 1.268\times10^{6}$, as shown by the solid red line in Figure \ref{fig1}. At the same time, changing the $\xi$ makes the $\tau_{\mathrm{D}}-\chi$ curve obtained from timing data intersect the lower bound of the black lines, we obtain the lower limit of the $\xi$ is $\xi\approx 2.393\times10^{5}$, as shown by the red dot line in Figure \ref{fig1}. Therefore, after considering the observation data of $\chi$, the limit on the number of precession cycles $\xi$ of PSR J2013+3845 is $2.393\times 10^{5}<\xi<1.268\times10^{6}$.

Based on the values of the obtained $\xi$, we get the $\dot{B_{\mathrm{d}}}-\chi$ curve under the assumption that the internal magnetic field of the star is toroidal-dominated (TD), as shown in Figure \ref{fig2}. From this figure, it can be seen that under the TD assumption, $\dot{B_{\mathrm{d}}}<0$ at the observed 
value of $\chi$, which consistent with the magnetic decay theory. Our results may provide a useful reference for investigating the viscous dissipation process and the related magnetic inclination angle evolution of pulsars, and may also have important significance for the study of continuous GWE from the stars.
\section{SUMMARY and outlook}\label{sec4}
In this short review, we reinvestigate the magnetic inclination angles of pulsars, as well as their evolutions. By following the ~\cite{Cheng2019}, we constrain the number of precession cycles for PSR J2013+3845, by using the brake index and magnetic inclination angle under the theoretical model considering magnetic dipole radiation, gravitational wave emission and free precession which is damped by internal dissipation. The evolutions of the inclination angle and magnetic field decay are considered in this model. The constraints on $\xi$ range from several times $10^{5}$ years to several times $10^{6}$ years. Then we use the $\xi$ value to obtain $\dot{B_{\mathrm{d}}}-\chi$ curve of this pulsar. The result shows that the internal magnetic field structure of PSR J2013+3845 is likely to be dominated by the toroidal component. In the future work, we will consider more braking mechanisms, for example, pulsar wind loss, quantum vacuum friction, etc., and choose suitable method to determine $\chi$, such as the relativistic RVM by considering the optical aberration effect, and the modified rotation vector model of magnetars.

It is expected that more large-scale radio telescopes, such as FAST and QiTai radio Telescope (QTT) under construction, will conduct more observations of pulsars to give more and better observation results, so as to study and determine the braking mechanisms and the evolution process of magnetic inclination angle of pulsars.

\section*{Acknowledgments}
This work was performed under the auspices of National Key Research and Development Program of China (No. 2022YFC2205202), Major Science and Technology Program of Xinjiang Uygur Autonomous Region through No. 2022A03013-1, and the National Natural Science Foundation of China grants 12288102, 12041304 and 11847102. 
\bibliography{Wiley-ASNA}
\end{document}